\documentclass[12pt]{iopart}
\usepackage{enumitem}
\usepackage{graphicx}
\usepackage[utf8]{inputenc}
\usepackage[svgnames]{xcolor} 
\usepackage{lipsum}
\usepackage[super,sort,compress]{cite}
\usepackage{iopams} 
\expandafter\let\csname equation*\endcsname\relax
\expandafter\let\csname endequation*\endcsname\relax
\usepackage{amsmath}
\usepackage{bm}
\usepackage{hyperref}
\usepackage[utf8]{inputenc}
\usepackage[svgnames]{xcolor} 
\usepackage{lipsum}

\begin{document}
\title{Mutually disjoint, maximally commuting set of physical observables for optimum state determination}

\author{H S Smitha Rao$^1$, Swarnamala Sirsi$^1$, 
Karthik Bharath$^2$}

\address{$^1$ Yuvaraja's college, University of Mysore, Mysuru, India }
\address{$^2$ University of Nottingham, Nottingham, U.K.}
\ead{smitharao@ycm.uni-mysore.ac.in}
\begin{abstract}
We consider the state determination problem using Mutually Unbiased Bases(MUBs). For spin-1, spin-3/2 and spin-2 systems, analogous to Pauli operators of spin-1/2 system, which are experimentally implementable and correspond to the optimum measurement in characterizing the density matrix, we describe a procedure to construct an orthonormal set of operators from MUBs. The constructed operators are maximally commuting, can be physically realized, and correspond to physical observables. The method of construction is general enough to allow for extensions to higher-dimensional spin systems and arbitrary density matrices in finite dimensions for which MUBs are known to exist. 
\end{abstract}

%
\vspace{2pc}
\noindent{\it Keywords}: Mutually Unbiased Bases (MUBs), Complete Set of Commuting Operators, optimum measurement. 
%
%
%
%

\section{Introduction}
The state of an $n$-dimensional system is represented by an $n \times n$ density operator, which needs $n^{2}-1$ real independent parameters for its complete specification. Any density operator in its eigenbasis can be represented as $\rho= \sum _{i=1}^{n} p_{i} \hat{P_{i}}$, where $p_{i}= Tr(\rho \hat{P}_{i})$ are the fractional populations. Thus in an eigenbasis, measurement of projection operators $\hat{P_{i}}$ yields $n-1$ independent probabilities corresponding to the diagonal elements of $\rho$. The remaining $n^2-1-(n-1)$ off-diagonal elements can be accessed by expressing $\rho$ in no fewer than $n+1$ bases. 
An optimal measurement strategy thus corresponds to a judicious choice of exactly $n+1$ bases, referred to as Mutually Unbiased Bases (MUBs)\cite{Schwing, woot}, such that measurement performed in each of these bases yields unique, non-redundant information about the  system.

For a given system, the practical utility of MUBs is dictated by their existence, and their physical realisation in a laboratory. The question of their existence for arbitrary $n$-dimensional systems has been extensively studied, and has been answered in the affirmative when $n$ is a prime or a power of a prime\cite{Ivano, woot}. For such spaces there always exist $n+1$ sets of MUBs which are complete. In particular when $n=2^{d}$ for some positive integer $d$, it is possible to find a partitioning of $d$-qubit Pauli operators into $n+1$ disjoint maximal commuting classes, where each class consists of $n-1$ maximally commuting set of operators \cite{band, zeii}. The corresponding MUBs are the simultaneous eigenbases of the $n+1$ commuting classes. More generally, in the case of prime power dimensions, several approaches are available to construct a complete set of $n+1$ MUBs (e.g. Heisenberg- Weyl group method\cite{band}; using finite field theory \cite{woot, tdurt} ; using generalized angular momentum operators\cite{kibler}). However, the existence of a complete set of MUBs for general finite-dimension Hilbert spaces remains an open question\cite{zan, gra, weig, beng}. 

When the $n+1$ sets of MUBs are known to exist, their construction and physical realization has been achieved for very specific applications such as quantum state tomography \cite{fili, fern, adam}, Mean's King problem \cite{eng, ara, hay}; quantum cryptography \cite{bou, lin}, quantum error correction \cite{cal}, entanglement detection \cite{ent}, and quantum coding and discrete Wigner function \cite{gib, bj}. Several experimental techniques to implement the complete set of MUBs in photonic systems have been investigated \cite{lima, adam}. 

Some recent works have focussed on generalising the notion of MUBs \cite{amir}. However, for systems for which they are known to exist, construction of optimal measurement operators based on MUBs is critical for their wider applicability. A general construction mechanism of such operators, to our knowledge, is unavailable in the literature. Our focus in this paper is to fill this void. Specifically, we consider spin systems for which MUBs are known to exist, and:
\begin{enumerate}
\item provide a general method to construct optimal measurement operators based on MUBs that are mutually disjoint and maximally commuting; 
\item identify the physical observables to which they correspond;
\item demonstrate how they can be physically realised.
\end{enumerate}

In order to concretize ideas we eschew a general exposition for arbitrary systems, and instead consider specific spin systems for which physically realizable, optimal measurement operators based on MUBs are hitherto unavailable. In particular, we construct an orthonormal set of operators based on MUBs for spin-1, spin-3/2 and spin-2 systems. We  achieve this based on a construction mechanism that extends the Stern-Gerlach setup for spin-1/2 systems. Such an extension enables us to identify the corresponding physical observables, as with the spin-1/2 case. We demonstrate how the operators can be physically realised for spin-1 and spin-3/2 cases. \emph{From an operational perspective, a key feature of the construction is that it naturally classifies the operators into mutually disjoint subsets, members of which commute enabling simultaneous measurements, resulting in a optimal measurement strategy}. The circumscription of the proposed methodology to spin systems is mainly in the interest of exposition. Examination of the method of construction will reveal that it is general enough to be applicable to higher-order spin systems and to arbitrary non-spin systems of finite dimension for which MUBs are known to exist. \\

\section{Spin-1/2 density matrix} 
Our technique is closely related to the case involving a spin-1/2 density matrix. It is instructive first to review this case along with the appropriate definitions. In a finite dimensional Hilbert space $H_{d}$, orthonormal bases $A= \{|a_{0}\rangle, |a_{1}\rangle,\ldots, |a_{d-1}\rangle\}$ and $B=\{|b_{0}\rangle, |b_{1}\rangle,\ldots, |b_{d-1}\rangle\}$ are said to be mutually unbiased if $|\langle a_{i}|b_{j}\rangle|= d^{-1/2}$, for every $i,j= 0,1,\ldots,d-1$.
For a spin-1/2 density matrix parameterized as $\rho=\frac{1}{2}(I_2+\sigma_xp_x+\sigma_yp_y+\sigma_zp_z)$, it is well known that the eigenbases of Pauli operators $\sigma_x, \sigma_y$ and $\sigma_z$ are the MUBs given by, 
\begin{align*}
B_{1}&= \{ |0 \rangle, |1 \rangle \},\\
B_{2}&=  \left\{ \frac{1}{\sqrt {2}} (|0 \rangle +  |1 \rangle ), \frac{1}{\sqrt {2}} (|0 \rangle -|1 \rangle )\right\},\\
B_{3}&= \left\{ \frac{1}{\sqrt {2}} (|0 \rangle + i |1 \rangle ), \frac{1}{\sqrt {2}} (|0 \rangle -i|1 \rangle )\right\},
\end{align*}
where $|0 \rangle= \left(
\begin{array}{c}
1\\
0\\
\end{array}
\right)$ and $|1\rangle= \left(
\begin{array}{c}
0\\
1\\
\end{array}
\right)$.\\ \\
Optimal measurement operators based on $B_i,i=1,2,3$  can be constructed and physically realized with the Stern--Gerlach apparatus. A detailed analysis of Stern--Gerlach experiment and its implications are extensively discussed in the literature \cite{swift, home, tekin} and references therein. In this experiment, a particle with magnetic moment $\vec{\mu}$ passes through the inhomogenous magnetic field $\vec{B}$. The potential energy associated with the particle is $\hat{\mathcal{H}}= - \vec{\mu}. \vec{B}$, where the magnetic moment  $\vec{\mu}$ is proportional to spin. Thus when the magnetic field is oriented along $z$- direction, one can measure the expectation value of $\sigma_{z}$. In terms of the projection operators $\hat{P_{1}}$ and ${\hat{P_{2}}}$ from the two eigenvectors of the operator $\sigma_z$, it is easy to see that $\sigma_{z}= \hat{P_{1}}- \hat{P_{2}}.$

The unitary matrix
{\small
\begin{align*}
U =	\frac{1}{\sqrt{2}}	 \left(
\begin{array}{cc}
  1 & 1 \\
1 & -1 \\
\end{array}\right),
\end{align*}
}
transforms $B_{1}$ to $B_{2}$. The observable $\sigma_x$ can be measured using the same apparatus if its diagonal (eigen) basis is of the same form as $\sigma_z$, necessitating that $\sigma_{x}= \hat{P'_{1}}-\hat{P'_{2}}$, where $\hat{P'_{i}}, i=1,2$ are the projection operators of $\sigma_x$. Experimentally this can be implemented by applying magnetic field in $x$-direction. In a similar manner from a unitary transformation of $B_1$ to $B_3$ we obtain $\sigma_{y}= \hat{P''_1}- \hat{P''_2}$ for appropriate projection operators, resulting in three measurements that constitute a complete set of parameters characterizing the spin-1/2 density matrix.

A key observation which we profitably exploit in the sequel is that the Pauli operators are linear combinations of projection matrices constructed from different basis vectors spanning a two-dimensional Hilbert space, related through unitary transformations that relate the basis sets of the MUBs. 
\section{Spin-1 density matrix}
Simultaneous measurement of a complete set of commuting operators is equivalent to the measurement of a single nondegenerate operator by means of a maximal or complete quantum test \cite{peres}. In some cases, generalization of Stern- Gerlach experiment for spin \textgreater 1/2 is possible by using electric multipole fields along with multipole magnetic fields \cite{lamb}. Measurements of spin-1 systems require an electric quadrupole field in addition to a dipole magnetic one \cite{swift}. 
To perform such measurements one requires four observables whose eigenstates are mutually unbiased; this, however, is not possible for spin components. Thus one cannot easily generalize the spin-1/2 Stern-Gerlach experiment as more number of parameters are needed. 

A spin-1 density matrix $\rho$ is characterized by eight independent parameters, and can be expanded using eight orthonormal (excluding identity) operators in infinitely many ways. Extending the program used for the spin-1/2 case requires a representation of the $3 \times 3$ density matrix $\rho$ in a matrix basis that mimics the role played by the Pauli operators. Since the natural choice for spin-$j$ Hamiltonian requires a mutlipole expansion, we choose the spherical tensor representation of the spin density matrix due to Fano \cite{fano}. The density matrix for any spin-$j$ system is given by, $\rho= \sum_{k,q} t^{k}_{q} {\tau^{k}_{q}}^{\dagger}$ where the irreducible spherical tensors $\tau^{k}_{q}$s are the $k^{th}$ degree polynomials constructed out of spin operators, $\vec{J}= (J_{x}, J_{y}, J_{z})$(See Appendix for the detailed description of the representation). For a spin-1/2 system, $\sigma_{z}= \tau^{1}_{0}$. As with SU(3) generators (for e.g. generalised Gell-Mann matrices), the spherical tensor operators for spin-1 density matrix consist of two diagonal matrices $\tau^1_0$ and $\tau^2_0$ which play the role of the diagonal $\sigma_z$.

In summary, optimal measurement of a spin-1 system can be achieved by an MUB consisting of four basis sets, each of which contains two commuting operators constructed using three projection operators. 
\subsection{Construction of maximally commuting orthogonal operators} 
From the discussion above, and guided by the fact that there are two diagonal matrices amongst the spherical tensors, in order to extend the technique from the spin-1/2 case,  we construct an orthonormal basis matrix set consisting of four sets of operators each containing two commuting operators, which enables simultaneous measurement using a single experimental setup.

Analogous to the spin-1/2 case where the Pauli operators $\sigma_x, \sigma_y$ and $\sigma_z$ are linear combinations of the projection operators, we define eight operators, comprising an orthonormal set, as linear combinations of the projection operators arising from the MUB. The coefficients of the linear combinations are chosen in a manner that ensures that the eight operators constitute the requisite maximally commuting orthogonal set. 

For a $3 \times 3$ spin-1 density matrix, from each of the four basis sets comprising an MUB $\{\{\psi_i\},\{\phi_j\}, \{\theta_k\}, \{\xi_l\}, i,j,k,l= 1, 2, 3\}$,  three projection operators can be constructed. In the canonical $|jm\rangle$ basis $\{\psi_1,\psi_2,\psi_3\}$ with projection operators $\hat{P}_i,i= 1, 2 , 3$, define first the commuting operators $\hat{\alpha_{1}}= \sum_{i} r_{i} \hat{P_{i}}$ and $\hat{\alpha_{2}}= \sum_{i} s_{i} \hat{P_{i}}$ for coefficient vectors $\vec{r}=(r_1,r_2,r_3)$ and $\vec{s}=(s_1,s_2,s_3)$. The two operators are orthogonal if $\vec{r}.\vec{s}=\sum_{i} r_{i}s_{i}= 0$, since
\begin{align*}
Tr(\hat{\alpha}_1\hat{\alpha}_2) &=
\sum_{i,j}r_{i} s_{j} Tr(\hat{P}_{i} \hat{P}_{j}) \\
&=\sum_{i,j} r_{i}s_{j} \delta_{ij}
=\sum_{i} r_{i}s_{i}.
\end{align*}
Furthermore, for the operators to be traceless, we require $\sum \limits_{i}r_{i}= \sum \limits_{i}s_{i}=0.$ 

Guided by the spin-1/2 case, we demand that the next set of operators has the same form as that of angular momentum basis $|jm \rangle $. That is, we impose the condition that $\hat{\alpha}_3$ and $\hat{\alpha}_4$ be defined with projection operators constructed with $\{\phi_i,i=1,2,3\}$ using the same coefficient vectors $\vec{r}$ and $\vec{s}$. Consequently, for a set $\hat{P}'_i,i=1,2,3$ of projection operators obtained from a second basis $\{\phi_1,\phi_2,\phi_3\}$, we define $\hat{\alpha_{3}}= \sum_{i} r_{i} \hat{P}'_{i}$ and $\hat{\alpha_{4}}= \sum_{i} s_{i} \hat{P'}_{i}$. Orthogonality requirement amongst the $\hat{\alpha}_i$ implies that the for a unitary $U$, 
\begin{align*}
Tr\big(\sum_{i} r_{i} \hat{P}_{i} \sum_{j} &r_{j} \hat{P}'_j\big)= 
Tr\big(\sum_{i,j} r_{i} r_{j}( \hat{P}_{i} U  \hat{P}_{j} U^{\dagger}\big) \\
&=Tr\big(\sum_{i,j} r_{i} r_{j}(|\psi_{i} \rangle \langle \psi_{i}| U |\psi_{j} \rangle \langle \psi_{j} | U^{\dagger}\big) \\
&=\sum_{i,j} r_{i} r_{j} u_{ij} u^{*}_{ij}\\
&=\sum_{i,j} r_{i} r_{j} \langle \psi_{i}|\phi_{j} \rangle \langle \phi_{i}|\psi_{j} \rangle=0 ,
\end{align*}
and thus $\sum_{i,j} r_{i} r_{j}=0$. In similar fashion we have $\sum_{i,j} s_{i} s_{j} =0$ and $\sum_{i,j} r_{i} s_{j}= 0$. Using the same coefficient vectors $\vec{r}$ and $\vec{s}$, we can continue in a similar manner to suitably define $\hat{\alpha}_5$ and $\hat{\alpha}_6$ using $\{\theta_k,k=1,2,3\}$, and $\hat{\alpha}_7$ and $\hat{\alpha}_8$ using $\{\xi_l,l=1,2,3\}$. \\

\subsection{Physical interpretation} 
The exact nature of the MUBs $\{\{\psi_i\},\{\phi_j\}, \{\theta_k\}, \{\xi_l\}, i, j, k, l=1,2,3\}$ were given by Bandyopadhyay et al. \cite{band}, and are of the form
{\small
\begin{align*}
B'_{1} =		 \left(
\begin{array}{ccc}
  1 & 0 & 0\\
0 & 1 & 0\\
0 & 0 & 1\\
\end{array}\right), 
\hspace{1mm}
B'_{2} =		 \frac{1}{\sqrt 3}\left(
\begin{array}{ccc}
  1 & 1 & 1 \\
1 & \omega^{2} & \omega\\
1 & \omega & \omega^{2} \\
\end{array}\right),  
\end{align*}
}
{\small
\begin{align*}
B'_{3} =		 \frac{1}{\sqrt 3}\left(
\begin{array}{ccc}
 1 & 1  & 1 \\
\omega & 1 & \omega^{2} \\
1 & \omega & \omega^{2} \\
\end{array}\right), 
\hspace{1mm}
B'_{4} =		 \frac{1}{\sqrt 3}\left(
\begin{array}{ccc}
 1 & 1 & 1 \\
\omega^{2} &  \omega & 1 \\
1 & \omega & \omega^{2} \\
\end{array}\right),
\end{align*}
}
where columns of $B'_1, B'_2, B'_3$ and $B'_4$ are $\{\psi_i\}, \{\phi_j\}, \{\theta_k\}$ and $\{\xi_l\}$ respectively, and $\omega= e^{2\pi i/3}$. 

We now identify the physical observables corresponding to the operators $\hat{\alpha}_i,i=1,\ldots,8$ and discuss their implementation. Consider first $\hat{\alpha}_1$ and $\hat{\alpha}_2$. Choosing one of $\vec{r}=(r_1,r_2,r_3)$ or $\vec{s}=(s_1,s_2,s_3)$ to be in the x-y plane (say $\vec{r}$), and from the conditions $\sum_{i} r_{i}= \sum_{i} s_{i}=0$ with $\sum_{i,j} r_{i} s_{j}= 0$, the choice of vectors $\vec{r}$ and $\vec{s}$ reduce to $\vec{r}= (r, 0, -r)$ and $\vec{s}= (s, -2s, s)$, up to an arbitrary permutation.

Examining the Fano representation of density matrix, we see that $\tau^{1}_{0}= \sqrt{\frac{3}{2}} J_{z}$ and $\tau^{2}_{0}= \frac{3J^{2}_{z}-J^{2}}{\sqrt{2}}$ and their expectation values determine two of the expansion coefficients of density matrix in this representation. Explicit forms of $\tau^{1}_{0}$ and $\tau^{2}_{0}$ are given by,  

{\small
\begin{align*}
\tau^{1}_{0}=& {\sqrt{\frac{3}{2}}} \left(\begin{array}{ccc}
1 & 0 & 0\\
0 & 0 & 0\\
0 & 0 & -1\\
\end{array} \right),
\hspace*{1mm}
\tau^{2}_{0}=  \frac{1}{\sqrt 2} \left(\begin{array}{ccc} 
1 & 0 & 0\\
0 & -2 & 0\\
0 & 0 & 1\\
\end{array} \right),
\end{align*}}

\noindent where expectation values of $ \tau^{1}_{0}$ and $ \tau^{2}_{0}$ are respectively associated with the first and second order moments of $J_{z}$, and hence constitute experimentally measurable parameters. Thus we choose $\hat{\alpha_{1}}$ to be $\tau^{1}_{0}$ and $\hat{\alpha_{2}}$ to be $\tau^{2}_{0}$. In other words, $\vec{r}={\sqrt{\frac{3}{2}}}(1, 0, -1)$ and $\vec{s}= \frac{1}{\sqrt{2}}(1, -2, 1)$. 

Now the first set of commuting operators in terms of projection operators associated with $B'_{1}$ is given by,
\begin{equation*}
   \hat{ {\alpha_{1}}}= {\sqrt{\frac{3}{2}}} (\hat{P_{1}}-\hat{P_{3}}), \ \ \ \hat{{\alpha_{2}}}= \frac{1}{\sqrt 2} (\hat{P_{1}}-2\hat{P_{2}}+\hat{P_{3}}) .
\end{equation*}
The bases $B'_{1}$ and $B'_{2}$ are connected by the Fourier transformation $U'$ \cite{pawel} given by,
{\small
\begin{align*}
U' =	\frac{1}{\sqrt{3}}	 \left(
\begin{array}{ccc}
  1 & 1 & 1\\
1 & \omega^{2} & \omega \\
1 & \omega & \omega^{2} \\
\end{array}\right).
\end{align*}
} 
\noindent
Thus $\hat{\alpha_{3}}$ and $\hat{\alpha_{4}}$ can be written as $\hat{\alpha_{3}}= U' \hat{\alpha_{1}} U'^{\dagger}$ and $\hat{\alpha_{4}}= U' \hat{\alpha_{2}} U'^{\dagger}$. 
In a similar manner, transition from $B'_{2}$ to $B'_{3}$ can be obtained by one-axis twisting, $e^{-iS^{2}_{z}t}$ \cite{kit} for $t=2\pi/3$ and from $B'_{2}$ to $B'_{4}$ for $t=4\pi/3$. 

The orthonormal set of commuting observables $\hat{\alpha_{i}}$, $i= 1, \ldots,8$ is given by,
\begin{align*}
\hat{\alpha_{1}}&= {\sqrt{\frac{3}{2}}}(\hat{P_{1}}-\hat{P_{3}}), \quad \hat{\alpha_{2}}= \frac{1}{\sqrt 2} (\hat{P_{1}}-2\hat{P_{2}}+\hat{P_{3}}) ,\\
\hat{\alpha_{3}}&= {\sqrt{\frac{3}{2}}}(\hat{P'_{1}}-\hat{P'_{3}}) , \quad \hat{\alpha_{4}}= \frac{1}{\sqrt 2} ( \hat{P'_{1}}-2\hat{P'_{2}}+\hat{P'_{3}}) ,\\
\hat{\alpha_{5}}&= {\sqrt{\frac{3}{2}}}(\hat{P''_{1}}-\hat{P''_{3}}) , \quad \hat{\alpha_{6}}= \frac{1}{\sqrt 2} ( \hat{P''_{1}}-2\hat{P''_{2}}+\hat{P''_{3}}) ,\\
\hat{\alpha_{7}}&= {\sqrt{\frac{3}{2}}}(\hat{P'''_{1}}-\hat{P'''_{3}}) , \quad \hat{\alpha_{8}}= \frac{1}{\sqrt 2} (\hat{P'''_{1}}-2\hat{P'''_{2}}+\hat{P'''_{3}}),
\end{align*}
where projection operators $\hat{P_{i}}, \hat{P'_{i}}, \hat{P''_{i}}, \hat{P'''_{i}}$ ( $i=1,2,3$) are respectively associated with the bases $B'_{1}, B'_{2}, B'_{3}, B'_{4}$. The new orthonormal operator basis is explicitly given by, 
{\small
\begin{align*}
\hat{{\alpha}_{1}}&=  {\sqrt{\frac{3}{2}}} \left(\begin{array}{ccc}
1 & 0 & 0\\
0 & 0 & 0\\
0 & 0 & -1\\
\end{array} \right),
\qquad
\hat{\alpha_{2}}=  \frac{1}{\sqrt 2} \left(\begin{array}{ccc} 
1 & 0 & 0\\
0 & -2 & 0\\
0 & 0 & 1\\
\end{array} \right),\\
\hat{\alpha_{3}}&= {\frac{1}{\sqrt{2}}} \left(\begin{array}{ccc} 
0 & -i\omega & i\omega^{2}\\
i\omega^{2} & 0 & -i\omega\\
-i\omega & i\omega^{2} & 0\\
\end{array} \right), 
\quad
\hat{\alpha_{4}}= \frac{1}{\sqrt 2}\left(\begin{array}{ccc} 
0 & -\omega & -\omega^{2}\\
-\omega^{2} & 0 & -\omega\\
-\omega & -\omega^{2} & 0\\
\end{array} \right),\\ 
\hat{\alpha_{5}}&={\frac{1}{\sqrt{2}}} \left(\begin{array}{ccc} 
0 & -i  & i\omega^{2} \\
i  & 0 & -i  \omega^{2} \\
-i \omega & i \omega & 0\\
\end{array} \right), 
\quad
\hat{\alpha_{6}}=\frac{1}{\sqrt 2} \left(\begin{array}{ccc} 
0 & -1 & -\omega^{2}\\
-1 & 0 & -\omega^{2} \\
-\omega & -\omega & 0\\
\end{array} \right),\\
\hat{\alpha_{7}}& ={\frac{1}{\sqrt{2}}} \left(\begin{array}{ccc} 
0 & -i \omega^{2} & i\omega^{2} \\
i \omega & 0 & -i\\
-i \omega & i  & 0\\
\end{array} \right),
\quad
\hat{\alpha_{8}}= {\frac{1}{\sqrt{2}}} \left(\begin{array}{ccc} 
0 & -\omega^{2} & -\omega^{2}\\
-\omega & 0 & -1\\
-\omega & -1 & 0\\
\end{array} \right).
\end{align*}
}
\noindent
The new operator basis provides an expansion of $\rho$:
\begin{equation*} 
    \rho= \frac{1}{3}(I + \sum \limits_{i=1}^{8} a_{i} \hat{\alpha_{i}}),
\end{equation*}
where $a_{i}= Tr(\rho \hat{\alpha_{i}})$. The expansion based on the operators constructed using the MUBs in a certain sense constitutes an optimal measurement strategy---complete state determination amounts to determining the $a_i$, which is optimally done using the maximally commuting orthogonal operators $\hat{\alpha}_i, i=1,\ldots,8$. 

\subsection{Physical realization} In addition to dipole magnetic field in the Stern- Gerlach apparatus, if one applies an external electric quadrupole field, the resulting Hamiltonian in the multipole expansion is given by, 
\begin{equation*}
    \hat{\mathcal{H}}= \sum \limits_{k=0}^{2} \sum \limits_{q= -k}^{+k} h^{k}_{q} {\tau^{k}_{q}}^{\dagger}.
    \end{equation*}
When the electric quadrupole field with asymmetry parameter $\eta=0$ is along the $z$-axis of the Principal Axes Frame(PAF) of the quadrupole tensor and the dipole magnetic field is oriented along the same $z$-axis \cite{swarna}, $\hat{\mathcal{H}}$ takes the form
\begin{equation*}
    \hat{\mathcal{H}}= \sum \limits_{k=0}^{2} h^{k}_{0} \tau^{k}_{0}.
\end{equation*}
In this experimental setup, one can measure the expectation values of $\hat{\alpha_{1}}$ and $\hat{\alpha_{2}}$. Implementation of unitary transformations namely Fourier transform and one-axis twisting in the lab leads to the measurement of all the observables $\hat{\alpha_{3}},\hat{\alpha}_4,\ldots,\hat{\alpha_{8}}$.\\
\section{Construction for a spin-3/2 system}
For a spin-3/2 system, the MUBs comprise five basis sets given by,
{\small
\begin{align*}
B''_{1} =		 \left(
\begin{array}{cccc}
  1 & 0 & 0 & 0 \\
0 & 1 & 0 & 0\\
0 & 0 & 1 & 0\\
0 & 0 & 0 & 1\\
\end{array}\right), 
\qquad
B''_{2} =		 \frac{1}{2} \left(
\begin{array}{cccc}
  1 & 1 & 1 &1  \\
1 & -1 & 1& -1 \\
1& 1& -1 &-1\\
1& -1& -1 & 1\\
\end{array}\right), 
\qquad
B''_{3} =		 \frac{1}{2}  \left(
\begin{array}{cccc}
  1 & 1 & 1 & 1\\
i & -i & i & -i \\
i & i & -i &-i \\
-1 & 1 & 1 & -1\\
\end{array}\right), 
\end{align*}
}
{\small
\begin{align*}
B''_{4} =		 \frac{1}{2}  \left(
\begin{array}{cccc}
  1 & 1&1 & 1\\
i & -i & i& -i \\
1 & 1 & -1 & -1\\
-i & i & i&-i\\
\end{array}\right), 
\qquad
B''_{5} =		 \frac{1}{2} \left(
\begin{array}{cccc}
  1 & 1 & 1 & 1 \\
1 & -1& 1& -1\\
i & i & -i & -i \\
-i & i & i & -i\\
\end{array}\right) .
\end{align*}
}
Thus we construct five sets of mutually disjoint, maximally commuting set of operators $\hat{\beta}_{i}$, $i= 1, 2, \ldots,15$. Fano expansion of spin-3/2 density matrix consists of three diagonal operators $\tau^{1}_{0}$, $\tau^{2}_{0}$ and $\tau^{3}_{0}$ in the $|3/2 \ m \rangle$ basis where $m= -3/2, \ldots,+3/2$. Along the lines of what was done for the spin-1 case, we choose $\hat{\beta_{1}}$, $\hat{\beta_{2}}$ and $\hat{\beta_{3}}$ to be $\tau^{1}_{0}$, $\tau^{2}_{0}$, $\tau^{3}_{0}$, given by, 
{\small
\begin{align*}
\tau^{1}_{0} = \hat{\beta_{1}} =	\frac{1}{\sqrt{5}}	 \left(
\begin{array}{cccc}
  3 & 0 & 0 & 0 \\
0 & 1 & 0 & 0\\
0 & 0 & -1 & 0\\
0 & 0 & 0 & -3\\ 
\end{array}\right),
\qquad
\tau^{2}_{0}= \hat{\beta_{2}} =		 \left(
\begin{array}{cccc}
 1 & 0 & 0 & 0 \\
0 & -1 & 0 & 0\\
0 & 0 & -1 & 0\\
0 & 0 & 0 & 1\\
\end{array}\right),
\end{align*}
}
{\small
\begin{align*}
\tau^{3}_{0}= \hat{\beta_{3}} =	\frac{1}{\sqrt{5}}	 \left(
\begin{array}{cccc}
  1 & 0 & 0 & 0 \\
0 & -3 & 0 & 0\\
0 & 0 & 3 & 0\\
0 & 0 & 0 & -1\\
\end{array}\right),
\end{align*}
}
\noindent
where $\tau^{3}_{0}= \frac{1}{3\sqrt{5}} [4J_{z}^{3}- (J_{z} J^{2}_{x}+ J^{2}_{x} J_{z}+  J_{x} J_{z} J_{x}) (J_{z} J^{2}_{y}+J^{2}_{y} J_{z}+J_{y} J_{z} J_{y})]$.
In terms of projection operators obtained from the canonical basis,
\begin{eqnarray*}
\hat{\beta_{1}}= 1/{\sqrt{5}} (3\hat{P_{1}}+\hat{P_{2}}-\hat{P_{3}}-3\hat{P_{4}}), \\
\hat{\beta_{2}}= \hat{P_{1}}-\hat{P_{2}}-\hat{P_{3}}+\hat{P_{4}}, \\
\hat{\beta_{3}}= 1/{\sqrt{5}} (\hat{P_{1}}-3\hat{P_{2}}+ 3\hat{P_{3}}-\hat{P_{4}}).
\end{eqnarray*}\\
Similarly remaining four sets of operators each containing three commuting operators are constructed from their respective projection operators employing the same linear combinations as above.\\
Thus spin-3/2 density matrix can be expanded in the new basis as
\begin{equation*} 
    \rho= \frac{1}{4}(I + \sum \limits_{i=1}^{15} b_{i} \hat{\beta_{i}}).
\end{equation*}
With the suitable application of quadrupole electric field, dipole and octopole magnetic field, one can obtain $\hat{\beta_{1}}$, $\hat{\beta_{2}}$ and $\hat{\beta_{3}}$. As the unitary transformations connecting different MUB sets are known, implementation of these transformations results in the measurement of rest of the observables. 

\section{Construction for a spin-2 system}
For spin-2 system, with $\omega=e^{2 \pi i/5}$, the six sets of MUBs are given by
{\small
\begin{align*}
B_{1} &=		 \frac{1}{\sqrt{5}}  \left(
\begin{array}{ccccc}
  1 & 0 & 0 & 0 & 0\\
0 & 1 & 0 & 0 & 0\\
0 & 0 & 1 & 0 & 0\\
0 & 0 & 0 & 1 & 0\\
0 & 0 & 0& 0 & 1\\
\end{array}\right),   
\qquad \quad 
B_{2} =		 \frac{1}{\sqrt{5}}  \left(
\begin{array}{ccccc}
  1 & 1 & 1 & 1 & 1 \\
1 & \omega & {\omega}^{2} & {\omega}^{3} & {\omega}^{4}\\
1 & {\omega}^{2} & {\omega}^{4} & {\omega}  & {\omega}^{3} \\
1 & {\omega}^{3} & {\omega} & {\omega}^{4} & {\omega}^{2} \\
1 & {\omega}^{4} & {\omega}^{3} &{\omega}^{2} &  {\omega} \\
\end{array}\right), 
\\
B_{3}& =		\frac{1}{\sqrt{5}}  \left(
\begin{array}{ccccc}
1 & 1 & 1 & 1 & 1 \\
\omega & \omega^{2} & \omega^{3} & \omega^{4}& 1\\
{\omega}^{4} & \omega& \omega^{3} & 1 & \omega^{2} \\
{\omega}^{4} & \omega^{2} & 1 & \omega^{3} & \omega \\
{\omega} & 1 & \omega^{4} & \omega^{3} & \omega^{2}\\
\end{array}\right),  
\quad
B_{4} =		 \frac{1}{\sqrt{5}}  \left(
\begin{array}{ccccc}
1 & 1& 1 & 1 & 1\\
\omega^{2} & \omega^{3}& \omega^{4} & 1 & \omega\\
{\omega}^{3} & 1&\omega^{2}  & \omega^{4} & \omega \\
{\omega}^{3} & \omega  & \omega^{4} & \omega^{2} & 1\\
{\omega}^{2} & \omega & 1 & \omega^{4} & \omega^{3}\\
\end{array}\right),   
\\
B_{5} &=		\frac{1}{\sqrt{5}}  \left(
\begin{array}{ccccc}
1 & 1 & 1& 1& 1\\
\omega^{3} & \omega^{4}& 1& \omega& \omega^{2} \\
{\omega}^{2} & \omega^{4}&\omega & \omega^{3}& 1\\
{\omega}^{2} & 1 & \omega^{3} & \omega & \omega^{4}\\
{\omega}^{3} &\omega^{2} & \omega & 1 & \omega^{4} \\
\end{array}\right),   
\quad
B_{6} =		\frac{1}{\sqrt{5}}  \left(
\begin{array}{ccccc}
1 & 1& 1& 1& 1 \\
\omega^{4} & 1 & {\omega} & {\omega}^{2}& {\omega}^{3}\\
{\omega}& {\omega}^{3} & 1 & {\omega}^{2}& {\omega}^{4}\\
{\omega}&{\omega}^{4} &{\omega}^{2} & 1 & {\omega}^{3}\\
{\omega}^{4}& {\omega}^{3}& {\omega}^{2}& {\omega} & 1\\
\end{array}\right).
\end{align*}
}
 In this case, the four Fano spherical tensors, in terms of the projection operators, are given by, 
\begin{eqnarray*}
\tau^{1}_{0}= \hat{\gamma_{1}}= {\sqrt{\frac{5}{4}}} (\hat{P_{1}}-\hat{P_{2}}+\hat{P_{4}}-\hat{P_{5}}), \\
\tau^{2}_{0}= \hat{\gamma_{2}}= {\frac{1}{\sqrt{2}}}(2\hat{P_{1}}+\hat{P_{2}}-2\hat{P_{4}}-\hat{P_{5}}), \\
\tau^{3}_{0}= \hat{\gamma_{3}}= {\frac{1}{\sqrt{2}}}(\hat{P_{1}}-2\hat{P_{2}}-\hat{P_{4}}+2\hat{P_{5}}), \\
\tau^{4}_{0}= \hat{\gamma_{4}}= {\frac{1}{2}} (\hat{P_{1}}+\hat{P_{2}}-4\hat{P_{3}}+\hat{P_{4}}+\hat{P_{5}})
\end{eqnarray*}
In similar fashion the remaining five sets of commuting operators can be obtained by operating the unitary transformations connecting MUBs in the angular momentum basis to rest of the MUBs. Consequently, spin-2 density matrix can now be expressed as
\begin{equation*} 
    \rho= \frac{1}{5}(I + \sum \limits_{i=1}^{24} c_{i} \hat{\gamma_{i}}).
\end{equation*}

\section{Concluding remarks}
 We have provided a mechanism to construct mutually disjoint, maximally commuting operators for dimensions where MUBs are known to exist. Since these operators are maximally commuting, measurements with them correspond to optimal determination of the parameters characterizing the density matrix of the state of a system. 
 Our construction rests on a key observation that the Pauli operators used in a Stern-Gerlach experiment for spin-1/2 particles are linear combinations of projection operators constructed from different MUBs. Leveraging this observation, we construct Pauli-like operators with eigenbases that are MUBs for spin-1, spin-3/2 and spin-2 systems. For prime and prime power dimensions $d$ (where $d= 2j+1$), using the fact that there always exists a complete set of $d+1$ MUBs, we have constructed $d+1$ sets of mutually disjoint operators, containing $d-1$ commuting operators in each set in the following manner: 
 \begin{enumerate}
\item  Consider the first set of MUBs as canonical basis. 
\item Consider an orthonormal set of $d^{2}$ matrices, with $d$ diagonal matrices which includes the identity $I$. 	
For example, if the angular momentum basis is used as the canonical basis, then the diagonal matrices can be identified as the Fano spherical tensor operators $\tau^{k}_{0}$s with matrix elements $\langle jm |\tau^{k}_{0} | jm \rangle= \sqrt{2k+1} C(jkj; m0m)$, where $k=0 \ldots d-1$ and $C(jkj; m0m)$ are the Clebsch-Gordan coefficients.
\item Express each diagonal matrix (excluding identity) as a linear combination of projection operators of the canonical basis. 
\item Identify the unitary transformations $U_{1}, U_{2}, \ldots, U_{d}$ that connect the first set of MUBs with the rest of MUB states. 
\item Implementing $U_{i} \tau^{k}_{0} U^{\dagger}_{i}$, where $i= 1, \ldots, d$ and $k=1 \ldots, d-1$, generate the complete set of mutually disjoint, maximally commuting set of operators. 
\end{enumerate}
Inspection of our method reveals that the main requirement for extension to higher-order spin systems and arbitrary density matrices representing non-spin systems is that the MUBs are known to exist and are available. For non-spin systems, expansion of the density matrix $\rho$ in an operator basis different from the spherical tensors can be considered. Physical realization then amounts to the identification of a suitable Hamiltonian that plays a role analogous to the multipole fields used in spin-$j$ systems.

Spin-5/2 is an intriguing case as it corresponds to the lowest composite dimension $d=6$ that is not a power of a prime for which the existence of a complete set of MUBs is yet to be established. It has been conjectured by Zauner \cite{zan} that for $d=6$ the maximum number of MUB sets is three. If the conjecture is true, then one cannot construct seven sets of mutually disjoint, maximally commuting set of operators. 

There have been numerous attempts to detect entanglement/correlation by using minimal number of experimentally viable local measurements. Recent works\cite{chris, bin} show that MUBs, as well as Mutually Unbiased Measurements(MUMs), can be efficiently used to detect entanglement in bipartite, multipartite and higher dimensional systems. In principle, our method of constructing mutually disjoint, maximally commuting set of operators may be harnessed to detect entanglement, since the eigenbases of our operators are MUBs. Since the constructed operators are maximally commuting, the detection mechanism would require a minimal number of measurements. This work will be taken up elsewhere.

\section*{Appendix}
The density matrix for a spin-$j$ system can be represented as
\begin{equation*}
\rho(\mathbf{J}):=\rho := \frac{1}{(2j+1)} \sum_{k=0}^{2j}\sum_{q=-k}^k t^k_q \tau^{k^\dagger}_q,
\end{equation*} 
$\tau^k_q$ are irreducible tensor operators of rank $k$ in the $2j+1$ dimensional spin space with projection $q$ along the axis of quantization in the real 3-dimensional space. The matrix elements of $\tau^{k}_{q}$ are 
\begin{equation*}
\langle jm' |\tau^{k}_{q}(\vec{J}) | jm \rangle= [k] C(jkj; mqm'),
\end{equation*} 
where $C(jkj;mqm')$ are the Clebsch--Gordan coefficients and $[k]=\sqrt{2k+1}$. $\tau^{k}_{q}$s satisfy orthogonality and symmetry relations,
\begin{equation*}
	Tr({\tau^{k^{\dagger}}_{q}\tau^{k^{'}}_{q^{'}}})= (2j+1)\,\delta_{kk^{'}} \delta_{qq^{'}}, \quad \tau^{k^{\dagger}}_{q} = (-1)^{q}\tau^{k}_{-q},
\end{equation*}
where the normalization has been chosen so as to be in agreement with Madison convention. 

The Fano statistical tensors or the spherical tensor parameters $t^k_q$ parametrize the density matrix $\rho$ as expectation values of $\tau^k_q$: $Tr(\rho\tau^k_q)=t^k_q$. Because of hermiticity of $\rho$, ${t^{k}_{q}}^{*}= (-1)^{q} t^{k}_{-q}$. 
The importance of irreducible spherical tensor operators lies in the fact they can be constructed as symmetrized products of the angular momentum operators $\mathbf{J}$ following  the well-known Weyl construction 
as,
$\tau_{q}^{k}(\mathbf{J}) =  \mathcal {N}_{kj}\,(\mathbf{J}\cdot \vec{\bf{\nabla}})^k \,r^{k} \,{Y}^{k}_{q}(\hat{r}),$
where 
\begin{equation*}
    \mathcal {N}_{kj}= {\frac{2^{k}}{k!}}{\sqrt{\frac{4\pi(2j-k)!(2j+1)}{(2j+k+1)!}}},
\end{equation*} 
are the normalization factors and ${Y}^{k}_{q}(\hat {r})$ are the spherical harmonics. The tensor operators are traceless but not Hermitian, and cannot in general be identified with generators of $SU(N)$. Also, the tensor operators $\tau^{k}_{0}$s are the physical observables which have the physical interpretation. That is, the  expectation values of $\tau^{k}_{0}$s correspond to the statistical moments and thus are measurable physical quantities. 
\section*{Acknowledgements}
HSS thanks the Department of Science and Technology(DST), India for the grant of INSPIRE Fellowship. KB's research is partially supported by NSF DMS 1613054 and NIH R01 CA214955.
\vspace*{-10pt}
\section*{References}
\bibliography{physicascripta}
\end{document}